\documentclass[]{article}
\usepackage{graphics}
\usepackage{graphicx}

\begin{document}

\newcommand{\be}{\begin{equation}}
\newcommand{\ee}[1]{\label{#1} \end{equation}}
\newcommand{\ba}{\begin{eqnarray}}
\newcommand{\ea}[1]{\label{#1} \end{eqnarray}}
\newcommand{\nl}{\nonumber \\}
\newcommand{\iint}{\int_0^{\infty}\!}
\newcommand{\pd}[2]{\frac{\partial #1}{\partial #2}}

\renewcommand{\d}[1]{{\rm d}#1}

\title{{\bf Higgs Pair Production at the LHC in Models with Universal Extra Dimensions}}

\author{
 {\sc H.~de Sandes} and {\sc R.~Rosenfeld} \\[1em]
 Instituto de F\'{\i}sica Te\'orica \\ State University of S\~ao Paulo, S\~ao Paulo, Brazil
}

\date{\today}

\maketitle

\begin{abstract}
In this letter we study the process of gluon fusion into a pair of Higgs
bosons in a model with one universal extra dimension.
We find that the contributions from the extra top quark Kaluza-Klein excitations 
lead to a
Higgs pair production cross section at the LHC that can be significantly
altered compared to the Standard Model value for small values of the
compactification scale.
\end{abstract}

\section{Introduction}

In spite of the great experimental successes of the Standard Model (SM), 
we still do not have a direct test of its symmetry breaking sector.
In fact, one may say that there is a tension arising from the indirect
bounds on the Higgs mass coming from the loop structure and precision
measurements, which favours a Higgs mass that is already excluded by 
direct searches.
In addition, we already know that the SM is incomplete, since it does not
provide for a viable dark matter candidate and for neutrino masses and mixings.

On a more theoretical side, the SM is not satisfactory due to the triviality and
hierarchy problems in the Higgs sector. These problems suggest that the Higgs sector 
should be viewed as an effective theory valid up to a certain energy scale.
The traditional solutions to these problems used to be represented by two broad classes
of models: supersymmetry and technicolor.

Recently a third class of solutions has been proposed, involving the existence
of extra space-like compact dimensions \cite{review}. This class can be further divided in three
different classes depending on the fields that can propagate in the extra dimensions and
the geometry of these extra dimensions. In this letter we will concentrate on the so-called
models of Universal Extra Dimensions (UED), where all fields can propagate in the flat compact 
extra dimensions \cite{UED}. 
An important property of UED arises from momentum conservation in the extra
dimensions. This implies that KK number is conserved in all tree level
vertices.
As a consequence contributions to electroweak observables arise only from loops of KK
particles allowing compactification scales as low as $500$ GeV \cite{ewpm}. 

We are interested in the consequences of this model for Higgs boson production
at hadron colliders. The analysis of single Higgs production in gluon
fusion process was done by Petriello \cite{petriello} and a significant enhancement
compared to the SM was found. 

Higgs pair production via gluon fusion in the SM was studied in \cite{higgspair}
and is an interesting process since it could give information on the Higgs boson
cubic coupling. Hence it is important to examine possible deviations from SM
predictions in different models. For instance, Higgs pair production in Little Higgs models
was studied in \cite{littleHiggs}.

In this paper we study the modifications of the Higgs pair production cross 
section via gluon fusion in UED.

\section{Model and relevant masses and couplings}

In models of UED all fields are allowed to propagate in the bulk
and hence they all have an associated Kaluza-Klein (KK) tower.
We will work in the case of one additional compact dimension.
In order to retain the zero modes corresponding to SM particles
it is usual to compactify the extra dimension in an orbifold
$S_1/Z_2$, defined by the identification $y \rightarrow y +  \pi R$,
where $y$ is the 5th dimension coordinate and $R$ is the compactification
radius, and demand that the fields with zero modes 
to be even under the transformation $y \rightarrow -y$.

After compactification the relevant fields for our purposes 
(Higgs doublet $H$, top quark singlet $t$, top quark doublet $Q$ and 
gluon field $G$) will have the usual KK expansion:
\begin{equation}
H(x^{\mu },y)=H^{0}(x)\chi ^{(0)}+\sum_{n=1}^{\infty }[H^{(n)}(x)\chi
^{(n)}(y)]
\end{equation}
\begin{equation}
t(x^{\mu },y)=t_{R}^{0}(x)\chi ^{(0)}+\sum_{n=1}^{\infty
}[t_{R}^{(n)}(x)\chi ^{(n)}(y)+t_{L}^{(n)}(x)\phi ^{(n)}(y)]
\end{equation}
\begin{equation}
G_{\mu }^{a}(x^{\nu },y)=G_{\mu }^{a (0)}(x)\chi
^{(0)}+\sum_{n=1}^{\infty }[G_{\mu }^{a (n)}(x)\chi ^{(n)}(y)]
\end{equation}
\begin{equation}
Q(x^{\mu },y)=Q_{L}^{0}(x)\chi ^{(0)}+\sum_{n=1}^{\infty
}[Q_{L}^{(n)}(x)\chi ^{(n)}(y)+Q_{R}^{(n)}(x)\phi ^{(n)}(y)]
\end{equation}
\begin{equation}
G_{5}^{a}(x^{\nu },y)=\sum_{n=1}^{\infty }[G_{5}^{a (n)}(x)\phi
^{(n)}(y)]
\end{equation}
where $\chi ^{(n)}(y)$ and $\phi ^{(n)}(y)$ are orthogonal basis:
\begin{equation}
\chi ^{(n)}(y)=\frac{1}{\sqrt{\pi R}}\cos \frac{ny}{R}, \qquad 
\chi ^{(0)}= \frac{1}{\sqrt{2\pi R}},
\qquad \phi ^{(n)}(y)=\frac{1}{\sqrt{\pi R}}\sin \frac{ny}{R}.
\end{equation}

The couplings of KK top quarks with gluons and Higgs field that enter 
in our computation are derived from the lagrangian:
\begin{equation}
{\cal L}_{top} = \int_{-\pi R}^{\pi R}dy\int d^{4}x\{i\overline{Q}D\!\!\!\!/Q+i\overline{t}%
D\!\!\!\!/t+[\lambda _{5}^{t}\overline{Q}i\sigma _{2}H^{\ast
}t+h.c.]\},
\end{equation}
where the covariant derivative is:
\begin{equation}
D\!\!\!\!/=\Gamma ^{M}(\partial _{M}-ig_{5}T^{a}G
_{M}^{a}),
\end{equation}
$M$ is a Lorentz index with values $M =0,1,2,3$ and
$5$ (we will use lower case greek index like $\mu=0,1,2,3$ to denote the
usual non-compact dimensions), 
$\Gamma ^{M}=(\gamma^{\mu},i\gamma ^{5})$  are the 5-d Dirac matrices,
$g_{5}$ and $\lambda _{5}^{t}$ are the 5-d QCD and Yukawa coupling constants respectively
and $T^{a}$ is the usual color group generator.
 
Considering only the coupling to the Higgs zero mode, which after spontaneous symmetry 
breaking is written in terms of its vacuum expectation value $v = 246$ GeV as
\begin{equation}
H^{(0)}=\frac{1}{\sqrt{2}}\left(
\begin{array}{c}
0 \\
v+h^{(0)}(x)
\end{array}
\right)
\end{equation}
 and using the relation between the 5-d and 4-d top Yukawa coupling,
$\lambda ^{t}=\frac{\lambda _{5}^{t}}{\sqrt{2\pi R}}$, with the zero mode
top quark mass given as usual by $m_{t}=\frac{\lambda^{t}v}{\sqrt{2}}$,
one finds that the mass eigenstates of the KK top quark tower are
\begin{equation}
T_{R}^{(n)}=\left(
\begin{array}{c}
t_{1R}^{(n)} \\
t_{2R}^{(n)}
\end{array}
\right) \qquad T_{L}^{(n)}=\left(
\begin{array}{c}
t_{1L}^{(n)} \\
t_{2L}^{(n)}
\end{array}
\right).
\end{equation}

These mass eigenstates are related to the original states by
\begin{equation}
T_{R}^{(n)}=U_{R}^{(n)}\left(
\begin{array}{c}
\tilde{t}_{R}^{(n)} \\
t_{R}^{(n)}
\end{array}
\right) \qquad T_{L}^{(n)}=U_{L}^{(n)}\left(
\begin{array}{c}
\tilde{t}_{L}^{(n)} \\
t_{L}^{(n)}
\end{array}
\right)
\end{equation}
where $\tilde{t}_{L}^{(n)}$ and $\tilde{t}_{R}^{(n)}$ denote the upper components of the 
doublets $Q_{L}^{(n)}$ and $Q_{R}^{(n)}$ respectively.
The orthogonal matrices 
$U_{R}^{(n)}$ and $U_{L}^{(n)}$  are given by
\begin{equation}
U_R^{(n)}=
\left(
\begin{array}{cc}
\cos \frac{\alpha ^{(n)}}{2} & \sin \frac{\alpha ^{(n)}}{2} \\
-\sin \frac{\alpha ^{(n)}}{2} & \cos \frac{\alpha ^{(n)}}{2}
\end{array}
\right) 
\qquad
U_L^{(n)}=
\left(
\begin{array}{cc}
\cos \frac{\alpha ^{(n)}}{2} & \sin \frac{\alpha ^{(n)}}{2} \\
\sin \frac{\alpha ^{(n)}}{2} & -\cos \frac{\alpha ^{(n)}}{2}
\end{array}
\right) 
\end{equation}
where $\sin \alpha^{(n)}\equiv \frac{m_{t}}{m_{t,n}}$ and 
$\cos \alpha ^{(n)}\equiv \frac{m_{n}}{m_{t,n}}$,
with $m_{t,n}^{2}\equiv m_{t}^{2}+m_{n}^{2}$ and
$m_{n}=\frac{n}{R}$.  
The two top KK towers mass eigenstates, $t_{1}^{(n)}$ and $t_{2}^{(n)}$, have a degenerate
mass given by $m_{t,n}$.

The couplings of the top KK tower states with gluons are simply given by
\begin{equation}
g_{s}{}\sum_{n=1}^{\infty }\left[\overline{t}_{1}^{(n)}g\!\!\!/^{(0)}%
t_{1}^{(n)}+\overline{t}_{2}^{(n)}g\!\!\!/^{(0)}t_{2}^{(n)} \right]
\end{equation}
whereas the coupling to the zero mode Higgs boson can mix $t_{1}^{(n)}$ and $t_{2}^{(n)}$:
 \begin{equation}
 \frac{m_{t}}{v}h^{(0)}{}\sum_{n=1}^{\infty }\left[\sin \alpha ^{(n)} \left(\overline{t}_{1L}^{(n)}t_{1R}^{(n)}+\overline{t}_{2L}^{(n)}t_{2R}^{(n)}\right)+\cos
\alpha^{(n)}\left(\overline{t}_{1L}^{(n)}t_{2R}^{(n)}-\overline{t}_{2L}^{(n)}t_{1R}^{(n)}\right) + h.c.\right]
\end{equation}
Notice that the top KK Yukawa couplings are proportional to the top quark mass and
hence their effects decouple for higher KK modes.

\section{Model implementation and results}

We implemented the new particles and couplings
in FeynArts \cite{feynarts} for an arbitrary number 
of KK modes. We then use FormCalc \cite{formcalc} to perform the
computation of traces and the reduction of the tensor one-loop integrals to
scalar Passarino-Veltman integrals \cite{pv}. Finally, LoopTools \cite{looptools}
computes numerically the integrals and CUBA \cite{cuba} integrates over phase space to
find the cross section. 
We verified that
in the case of single Higgs production in UED,
where only a triangle diagram contributes,
the program reproduces both analytically and numerically 
the results obtained by Petriello \cite{petriello}.
We have also checked our code with the SM Higgs pair
production \cite{higgspair}.

\begin{figure}[h,t,b]
\includegraphics[width=1\textwidth,angle=0]{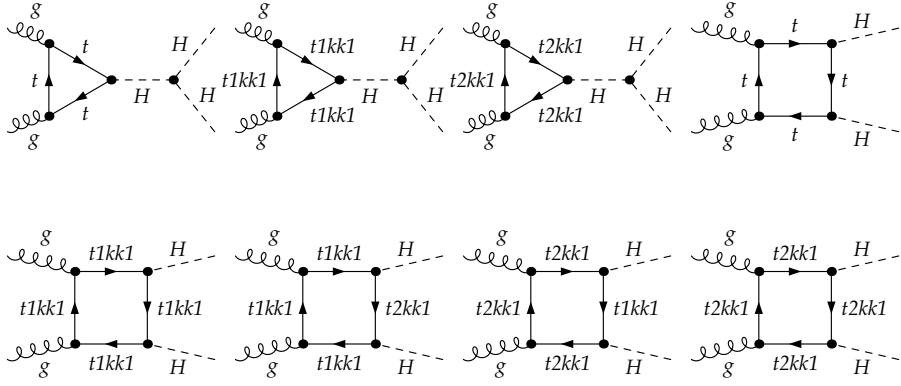}
\vspace{-7cm}
\caption{\label{diagrams}
Feynman diagrams for the process $gg \rightarrow HH$ with the contribution from the 
first top quark KK modes, denoted by $t1kk1$ and $t2kk1$ in the figure.
Permutations of the external lines are not shown.
}
\end{figure}

In Figure \ref{diagrams} we show the diagrams that are
computed for one top quark KK level. Notice the presence of 2
top KK excitations for each level and their mixture through 
the Yukawa coupling.

In this work we will consider compactification scales $1/R$ as low as $500$ GeV, as 
allowed by 
electroweak precision measurements \cite{ewpm} (see also \cite{bsgamma} for bounds
coming from $b \rightarrow s \gamma$ processes) and first compute the deviations
from the SM for a fixed partonic center-of-mass energy $\sqrt{\hat{s}}$ as a function of the Higgs boson mass $M_H$ for
different values of the compactification scale.
We will include in the calculation a number $n$ of KK levels such that $m_n < 10$ TeV for a given 
compactification scale, where one expects that 
the 4-dimensional effective theory starts to loose its validity \cite{UED}. For instance, we considered 
the contribution of $20$ KK levels for $1/R = 500$ GeV.
In practice, the convergence for large values of $n$ is very rapid.

\begin{figure}[h,t,b]
\includegraphics[width=1\textwidth,angle=0]{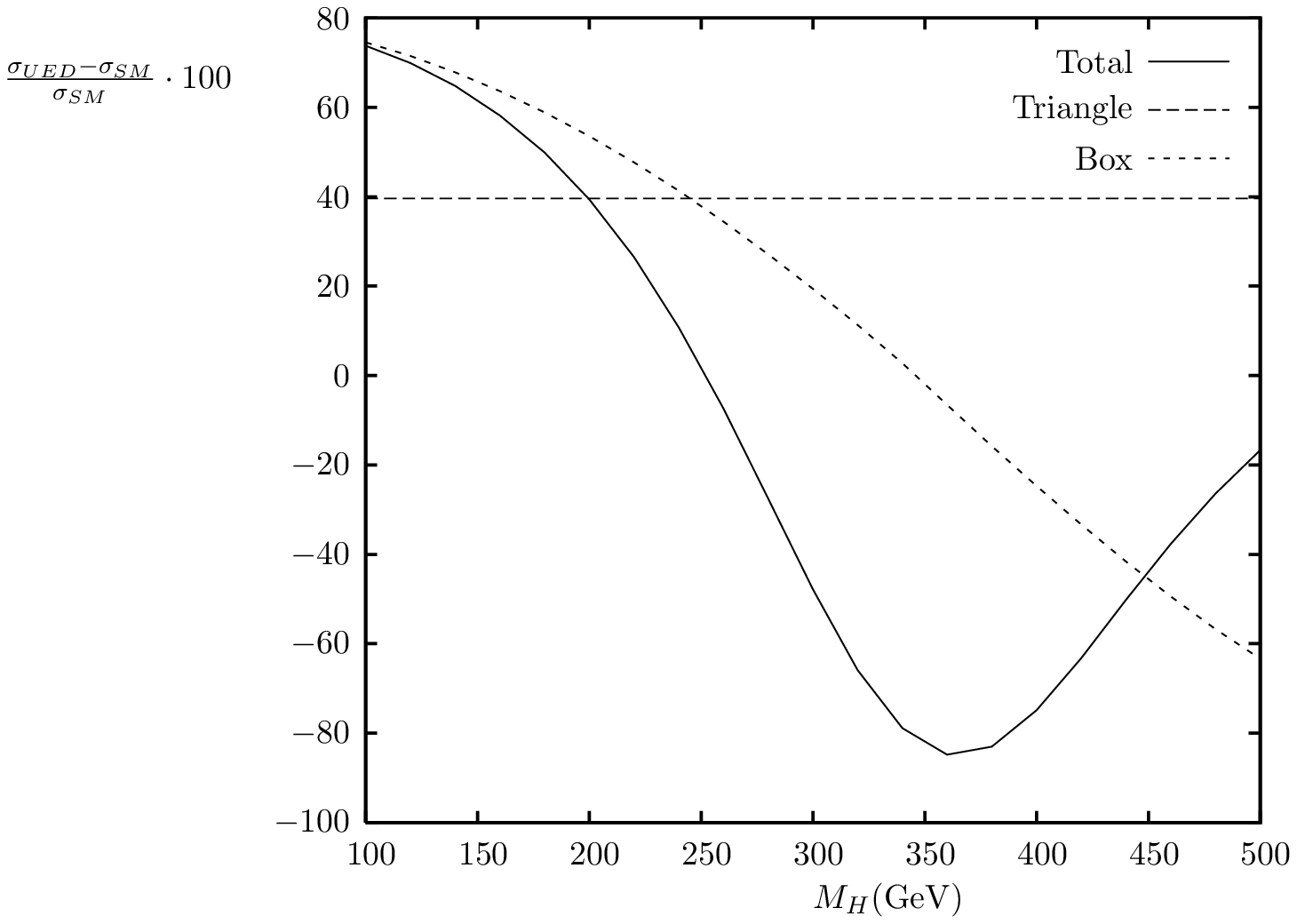}
\caption{\label{separate}
Deviations from SM arising separately from triangle and box contributions, together with the total deviation, as a function of
the Higgs boson mass. Center-of-mass energy is fixed at $\sqrt{\hat{s}} = 1$ TeV and compactification scale $1/R= 500$ GeV.
}
\end{figure}

In Figure \ref{separate} we show the differences between the SM and UED contributions for 
the triangle and box diagrams separetely. We fix $\sqrt{\hat{s}} = 1.0$ TeV and $1/R = 500$ GeV
for illustration.

\begin{figure}[h,t,b]
\includegraphics[width=1\textwidth,angle=0]{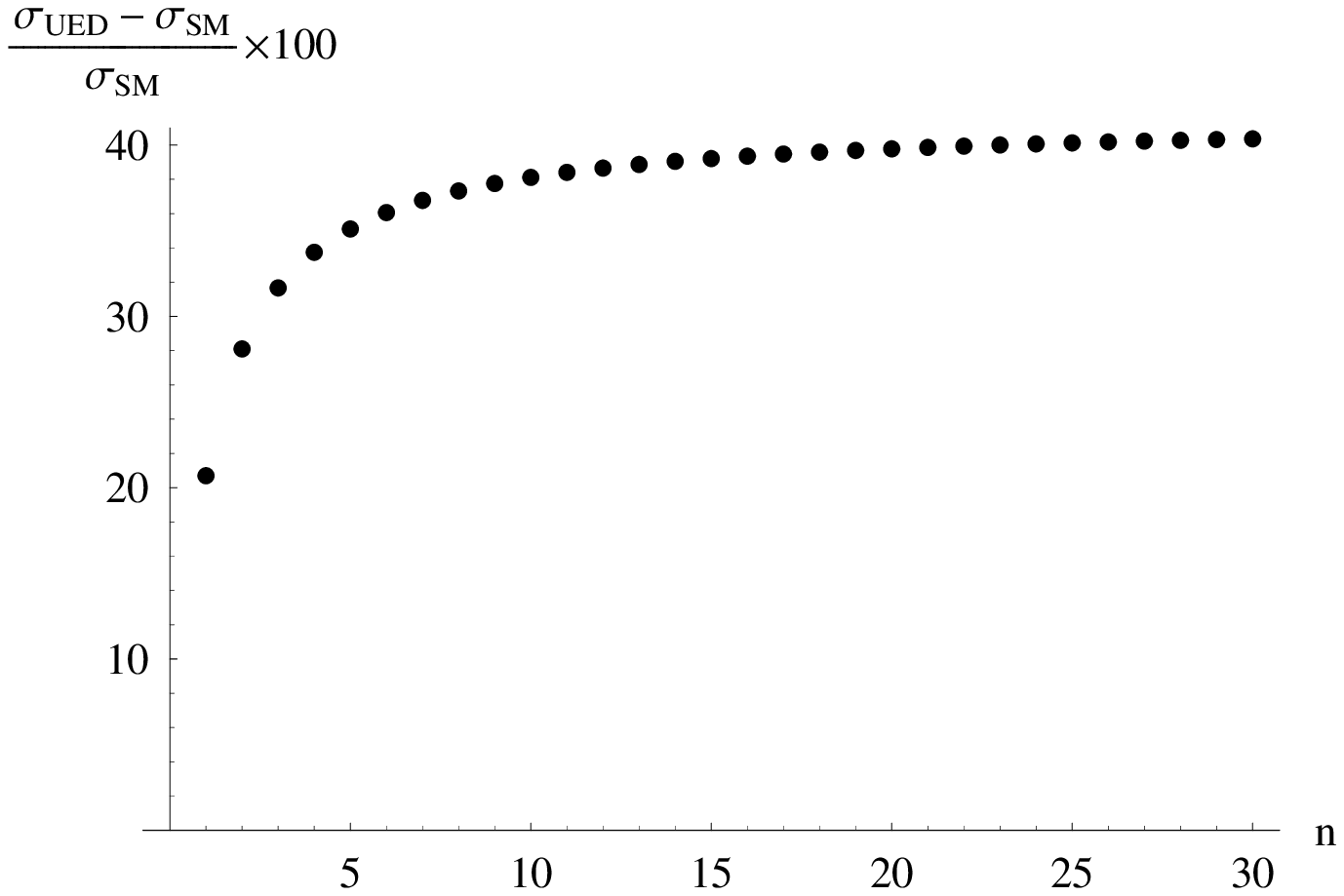}
\caption{\label{triangle}
Analytical calculation of deviations from SM arising only from the triangle as a function of the KK level.
Center-of-mass energy is fixed at $\sqrt{\hat{s}} = 1$ TeV and compactification scale $1/R= 500$ GeV.
}
\end{figure}

The triangle contribution can be understood analytically; it is given by the difference in the triangle amplitude:
\begin{equation}
\frac{\sigma_{UED}^{\triangle}(gg\rightarrow HH) - \sigma_{SM}^{\triangle}(gg\rightarrow HH)}
{\sigma_{SM}^{\triangle}(gg\rightarrow HH)} = \frac{ | A_{SM} + A_{KK}|^2 -  | A_{SM}|^2}{ | A_{SM} |^2},
\end{equation}
where
\begin{equation}
A_{SM} = m_t^2 \left[
(\hat{s}-4m_{t}^{2}) C_{0}(\hat{s},m_{t}^{2})-2\right],
\end{equation}
\begin{equation}
A_{KK} =2 m_t \sum_n m_{t,n}\sin \alpha^{(n)} \left[
(\hat{s}-4m_{t,n}^{2}) C_{0}(\hat{s},m_{t,n}^{2})-2\right],
\label{AKK}
\end{equation}
and, as usual
\begin{eqnarray}
C_{0}(\hat{s},m^{2}) =&  -\frac{2}{\hat{s}} \left[\arcsin\left(1/\sqrt{\tau} \right) \right]^2 \;\; & \mbox{for} \;\; \tau \ge1 \nonumber \\
    = &\frac{1}{2 \hat{s}} \left[ \log\left( \frac{1+\sqrt{1-\tau}}{1-\sqrt{1-\tau}} \right) - i \pi \right]^2 &  \mbox{for} \;\; \tau < 1
\end{eqnarray}
with $\tau = 4 m^2/\hat{s}$. Notice that the triangle contribution is independent of the Higgs boson mass.
The factor of $2$ in eq. (\ref{AKK}) is due to the presence of 2
top KK excitations for each level.
 Figure \ref{triangle} shows the analytical result for difference in the triangle contribution only, showing the rapid
convergence of the result and its agreement with the numerical computation 
shown in Figure \ref{separate}.

The box contribution is more difficult to analyze due to the fact that many Passarino-Veltman integrals with different
arguments appear in the result. This is the reason of the more complicated behavior of the box contribution
depicted in Figure \ref{separate}.
We notice that there is a strong interference between the triangle and box contributions. The final result shows large deviations
both enhancing and suppressing the cross section, depending on the Higgs boson mass. In the case of a Higgs boson lighter than $M_H = 200$ GeV, the partonic gluon fusion cross section can be enhanced by more than $40 \%$.  These deviations
increase with partonic center-of-mass energy.

\begin{figure}[h,t,b]
\includegraphics[width=1\textwidth,angle=0]{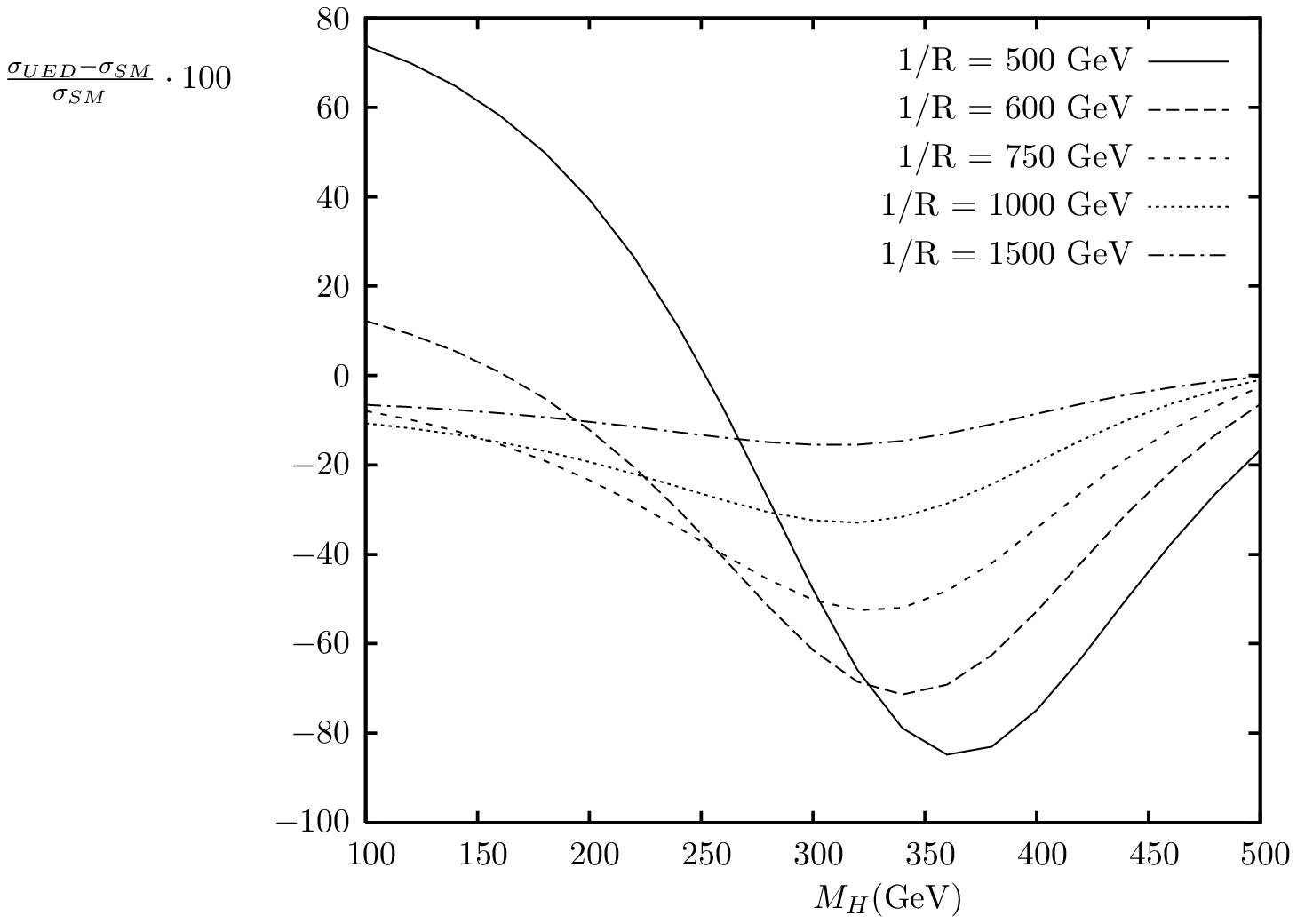}
\caption{\label{partonic}
Deviations from SM gluon fusion Higgs pair production arising from top KK modes as a function of
the Higgs boson mass for different compactification scales. Center-of-mass energy 
is fixed at $\sqrt{\hat{s}} = 1$ TeV.
}
\end{figure}

We present in Figure \ref{partonic} the deviations from the SM result
for the partonic gluon fusion Higgs pair production for different values of 
the compactification scale for a fixed value of the center-of-mass energy 
at $\sqrt{\hat{s}} = 1$ TeV.  As expected, for larger values
of $1/R$ the KK modes get heavier and the deviations from the SM rapidly decreases.

\begin{figure}[h,t,b]
\includegraphics[width=1\textwidth,angle=0]{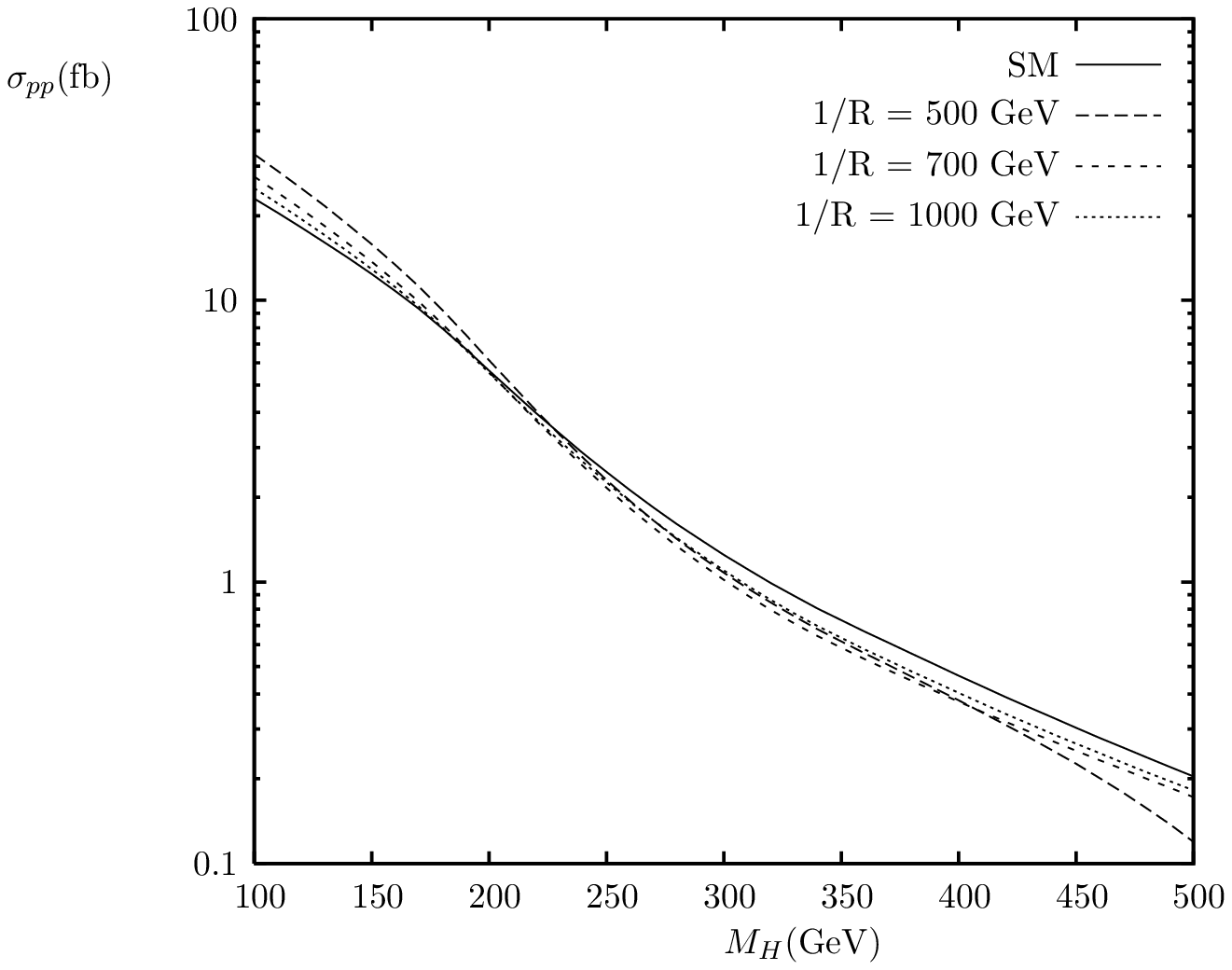}
\caption{\label{total}
SM Higgs pair production cross section via gluon fusion at the LHC 
as a function 
of the Higgs mass compared with the UED result for 
compactification scales of $1/R = 500, 700$ and $1000$ GeV.  
}
\end{figure}

\begin{figure}[h,t,b]
\includegraphics[width=1\textwidth,angle=0]{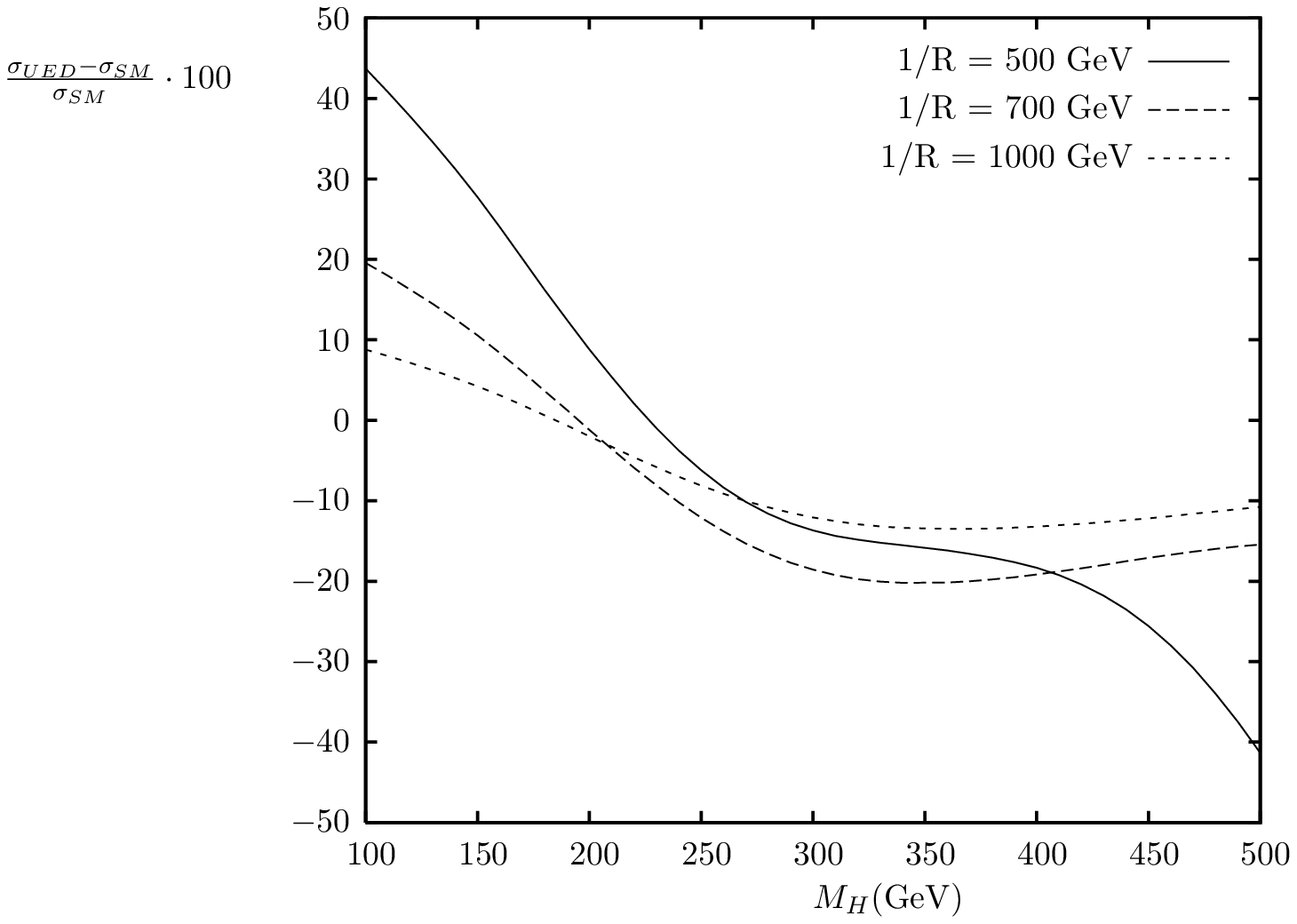}
\caption{\label{desvio}
Deviations from SM of Higgs pair production cross section via gluon fusion 
at the LHC as a function of the Higgs mass for values of the compactification
scales of $1/R = 500, 700$ and $1000$ GeV.
}
\end{figure}

The total Higgs pair production cross section at the LHC is computed in the standard way
by convoluting the partonic cross section with the gluon distribution function.
We used the Mathematica package implementation of the parton distribution functions of
\cite{alekhin} with factorization and renormalization scales given by $Q^2 = \hat{s}$, 
$\alpha_s(M_Z) = 0.118$ and $m_t = 175$ GeV.
In Figures \ref{total} and \ref{desvio} we compare the SM result as a 
function of the Higgs mass 
with the UED results with 
compactification scales of $1/R = 500, 700$ and $1000$ GeV.  
Differences as large as $\pm 40$\% can arise in these models. 

Electroweak precision data puts bounds on the compactification scale as a
function of the Higgs mass and these constraints decrease with
increasing Higgs mass \cite{ewpm}. 
These constraints are sensitive to the top mass allowing
$\frac{1}{R} = 600$ GeV for $m_H = 115$ GeV and $m_t =
173$ GeV , which increases in 23\% the SM cross section.
Bounds coming from $b \rightarrow s \gamma$ process \cite{bsgamma}
implies a compactification scale as low as $600$ GeV independent of the Higgs
mass. The cross section is increased by 16\% for a
light Higgs with mass $120$ GeV and $\frac{1}{R} = 700$ GeV.

\section{Conclusions}

In this paper we studied the effects of UED in the gluon fusion Higgs pair production 
cross section.
We implemented the contributions of the top KK excitations for the triangle and 
box diagrams and showed that the partonic cross section shows large deviations
both enhancing and suppressing the cross section, depending on the Higgs boson mass.
The total gluon fusion Higgs pair production cross section at the LHC can be 
modified by up to  $23$\% when bounds from precision measurements are taken into account.
These effects are rapidly reduced for larger values of the compactification scale.

\section*{Acknowledgments}
The work of H.~de Sandes is funded by a FAPESP doctoral fellowship.
R.~Rosenfeld thanks CNPq for partial financial support. 
We thank Claudio Dib and Alfonso Zerwekh for participation in the 
early stages of this project.

\end{document}